\documentclass[sigconf]{acmart}
\usepackage{multirow}

\AtBeginDocument{%
  }

\setcopyright{acmlicensed}
\copyrightyear{2024}
\acmYear{2024}
\acmDOI{XXXXXXX.XXXXXXX}

\acmConference[TSMO KDD'24]{Workshop on Two-sided Marketplace Optimization: Search, Pricing, Matching \& Growth}{August, 2024}{Barcelona, Spain}

\begin{document}

\title{Strategic Coupon Allocation for Increasing Providers'\\ Sales Experiences in Two-sided Marketplaces}


\author{Koya Ohashi}
\email{k-ohashi@mercari.com}
\affiliation{%
  \institution{Mercari, Inc.}
  \state{Tokyo}
  \country{Japan}
}

\author{Sho Sekine}
\email{s-sekine@mercari.com}
\affiliation{%
  \institution{Mercari, Inc.}
  \state{Tokyo}
  \country{Japan}
}

\author{Deddy Jobson}
\email{deddy@mercari.com}
\affiliation{%
  \institution{Mercari, Inc.}
  \state{Tokyo}
  \country{Japan}
}

\author{Jie Yang}
\email{j-yang@mercari.com}
\affiliation{%
  \institution{Mercari, Inc.}
  \state{Tokyo}
  \country{Japan}
}

\author{Naoki Nishimura}
\email{nishimura@r.recruit.co.jp}
\affiliation{%
  \institution{Recruit Co., Ltd}
  \state{Tokyo}
  \country{Japan}
}
\affiliation{%
 \institution{University of Tsukuba}
  \state{Ibaraki}
  \country{Japan}
}

\author{Noriyoshi Sukegawa}
\email{sukegawa@hosei.ac.jp}
\affiliation{%
  \institution{Hosei University}
  \state{Tokyo}
  \country{Japan}
}

\author{Yuichi Takano}
\email{ytakano@sk.tsukuba.ac.jp}
\affiliation{%
  \institution{University of Tsukuba}
  \state{Ibaraki}
  \country{Japan}
}

\renewcommand{\shortauthors}{Ohashi et al.}

\begin{abstract}
In a two-sided marketplace, network effects are crucial for competitiveness, and platforms need to retain users through advanced customer relationship management as much as possible. Maintaining numerous providers' stable and active presence on the platform is highly important to enhance the marketplace's scale and diversity. The strongest motivation for providers to continue using the platform is to realize actual profits through sales. Then, we propose a personalized promotion to increase the number of successful providers with sales experiences on the platform. The main contributions of our research are twofold. First, we introduce a new perspective in provider management with the distribution of successful sales experiences. Second, we propose a personalized promotion optimization method to maximize the number of providers' sales experiences. By utilizing this approach, we ensure equal opportunities for providers to experience sales without being monopolized by a few providers. Through experiments using actual data on coupon distribution, we confirm that our method enables the implementation of coupon allocation strategies that significantly increase the total number of providers having sales experiences. 
\end{abstract}

\ccsdesc[500]{Applied computing~Marketing}
\ccsdesc[500]{Applied computing~Online shopping}

\keywords{two-sided marketplace, uplift model, fairness, coupon allocation, customer relationship management, provider management}

\maketitle

\section{Introduction}
In the evolving landscape of two-sided marketplaces, the dynamics between supply side (e.g., listers in Airbnb, drivers in Uber, providers in flea market apps) and demand side (e.g., travelers, customers, consumers) play a pivotal role in shaping the competitive advantage and sustainability of the platform. Unlike traditional business-to-consumer (B2C) e-commerce services, two-sided marketplaces have a unique challenge: the platform cannot directly control the supply side, but it must rely on the voluntary participation of individual providers. 

The success of two-sided marketplaces depends significantly on the platform's ability to utilize and amplify network effects, where the value of the service increases as more participants join the network  \cite{Bruno_2021}. This characteristic necessitates innovative customer relationship management (CRM) to enhance provider lifetime value (LTV), thereby ensuring a steady and diversified supply to meet the growing demands of the marketplace. The CRM strategies differ from service to service. In this paper, we especially discuss services where actual product sales occur, such as online flea market apps (e.g., eBay, Etsy), rather than other two-sided market services, such as social network services.

There are various strategies to increase provider LTV \cite{zhou2020users}, such as providing tools that enhance their business operations, offering insights to optimize their sales, or creating loyalty programs that increase their attachment to the platform. However, no one doubts that deriving benefits from sales is the most fundamental value proposition influencing provider's satisfaction and, consequently, their retention and LTV. Therefore, it is necessary to distribute sales experiences to as many providers as possible to retain them. In other words, reducing the number of providers with zero sales experience is essential. In this paper, we define providers with sales experience, even once, as \textit{successful providers}. However, sales tend to be monopolized by a small number of providers without appropriate intervention from platforms \cite{DBLP:journals/corr/abs-2106-02702,10.1257/app.20160213, 10.1145/2998181.2998327}. For example, on flea market services, sales often cluster around a small group of super-heavy providers or highly experienced providers who are good at earning consumer views and impressions. This can lead to decreased sales opportunities, especially for new entrants, less experienced providers, or providers of niche items. In the case of a flea market app, each provider usually holds multiple items, and the number of items held by each provider varies significantly. Therefore, effective sales control through appropriate intervention by the platformer is crucial from a CRM perspective.

The concept of fairness has been actively researched as a solution to such unfair situations \cite{DBLP:journals/corr/abs-2002-10764, DBLP:journals/corr/Burke17aa,10.1145/3219819.3220088, Wang_2023, DBLP:journals/corr/abs-1805-01788}. Most research focuses on ensuring fairness in rankings in recommendation systems at the item level, aiming to achieve fairness in exposure chances for each item to equalize opportunities to get sold. However, these approaches are insufficient from two perspectives to enhance the number of successful providers.

First, fairness at the item level is not always equal to fairness at the provider level \cite{DBLP:journals/corr/Burke17aa,DBLP:journals/corr/abs-1907-13158, Wang_2023}. Particularly in flea market apps where individuals can easily list everyday items for sale, some providers can list hundreds of items while others list only a few items. Therefore, even if exposure opportunities are distributed fairly at the item level, there is a risk that sales will concentrate on a few providers who have listed many items. The item-level fairness approach is insufficient because we aim to maximize the number of successful providers.

Second, the distribution of exposure opportunities through recommendation rankings does not always lead to the fairness of outcomes \cite{Wang_2023}. Practically, in flea market apps, providers' skills vary, and simply increasing views of items may not always lead to successful sales due to factors such as incorrect pricing. Our goal is not just the fairness of opportunities for each provider but the fairness of outcomes by maximizing the number of successful providers. For promoting sales of such items, online/electronic coupons are considered one of the most effective marketing tools \cite{doi:10.1177/1356766712471839,DUAN2022102846}. To pursue the number of successful providers, we need to consider exposure opportunities and incentive interventions such as coupons.

{\bf Our Contributions:} We propose a coupon allocation method that maximizes the number of successful providers. Our contributions are as follows:

\begin{enumerate}
    \item We propose a new evaluation metric expressing the provider's sales experience probability. By maximizing this metric, we can aim to maximize the number of successful providers in the marketplace as a whole.

    \item We formulate a coupon allocation optimization problem by an uplift modeling framework with the evaluation metric, an integer nonlinear programming problem.

    \item We derive an integer linear programming formulation by linearizing the objective function of the nonlinear coupon allocation problem to make it usable in real-world scenarios where commercial nonlinear solvers are unavailable.

    \item We demonstrate that our proposed method achieves the best performance regarding the number of successful providers using a real-world dataset of coupon distribution. Consistent results are observed in multiple coupon distribution experiment rounds, indicating the performance's robustness.
\end{enumerate}

\section{Related work}
\subsection{Coupon allocation}
Albert and Goldenberg \cite{10.1145/3511808.3557100} utilized uplift modeling estimations for personalized promotions and discount optimization in an online e-commerce setting. They focused on purchase completion and incremental revenue as objective values and solved the optimization problem as a multiple-choice knapsack problem. While other studies have also addressed personalized promotions and discount optimization using uplift modeling \cite{zhao2019uplift,Baier2021Profit,uehara2024robust}, their emphasis is on maximizing short-term metrics like purchase completion during the promotion period. Retaining users in the long term is essential for maintaining a competitive advantage in a two-sided marketplace.

Yang, Li, and Jobson \cite{yang2022personalized} proposed personalized promotions optimization with direct and enduring effect prediction to evaluate the promotional effect even after the promotion period. They enhanced direct and enduring response training by applying a multi-task learning framework. However, they formulated promotion optimization for consumers, which can not directly apply to CRM for providers who usually have multiple items in our two-sided marketplace scenario. 

We propose personalized promotions optimization for providers using an uplift modeling framework in two-sided marketplaces. By setting the probability of provider sales experiences as the target variable, we aim to offer a value proposition to a larger number of providers as a platform, ultimately leading to medium to long-term retention of providers.

\subsection{Fair recommendation}
Fairness can vary in definition depending on the characteristics of the service and the stakeholders involved. Wang et al. \cite{Wang_2023} provided an overview of recent studies and gave an organized classification of fairness. Fairness can be distinguished based on the means and objectives. In terms of means, it is divided into process fairness and outcome fairness. Regarding objectives, it depends on which fairness of the subject is to be guaranteed. The problem we are addressing in this paper focuses on outcome fairness for providers. While there are several studies on item fairness in two-sided marketplaces \cite{DBLP:journals/corr/abs-1805-01788,10.1145/3340531.3411962}, there are few prior studies addressing provider fairness. Burke \cite{DBLP:journals/corr/Burke17aa} and Abdollahpouri and Burke \cite{DBLP:journals/corr/abs-1907-13158} categorized the provider fairness as the item fairness at the group level. However, they did not provide specific formulations of provider fairness and evaluations.

Fairness in recommender systems has been actively researched these days. Singh and Joachims \cite{10.1145/3219819.3220088} proposed a mathematical optimization framework to achieve fair item exposure allocation in rankings. Biega, Gummadi, and Weikum \cite{DBLP:journals/corr/abs-1805-01788} proposed the ranking system to ensure fair opportunity allocation, considering rankings significantly influence consumer attention towards items. However, in both cases, the discussions are around fairness at the item or consumer level rather than on the provider level. Additionally, simply increasing the visibility of items may not be sufficient, especially in online flea market services where item qualities vary widely. Our goal extends beyond achieving visibility fairness in recommendation systems to ensuring fairness in success experiences through promotions, with practical implications for platforms.

Saito and Joachims \cite{Saito_2022} aimed to achieve outcome fairness by maximizing the product of all expected revenues as the objective function. This objective function, possessing the characteristics of Nash Social Welfare (NSW), allows for maximizing impact while ensuring fairness. However, this NSW method can not be directly applied to provider fairness, although it is applicable to achieve item-level fairness because it achieves fairness through maximizing the product of all revenues, not the number of successful providers.

Then, we need to define a new evaluation metric for fairness at the provider level and further formalize it as an optimization problem for coupon allocation. We define a new provider fairness metric as the number of successful providers. To the best of our knowledge, no prior studies address provider fairness in the context of coupon allocation optimization using uplift modeling. This paper is the first to propose a coupon allocation strategy to maximize the total number of successful providers through the distribution of sales experiences, introducing a novel approach to fairness in this domain.

\section{Proposed model}
This section explains how to estimate the coupon effect and introduces the new evaluation metric of the provider's sales experience. We then formulate the coupon allocation optimization problem as an integer nonlinear programming problem and provide its linear transformation.

\subsection{Coupon effect estimation}
Uplift modeling is a predictive modeling approach designed to forecast the net impact of a specific action on a particular outcome \cite{doi:10.1089/big.2017.0104}. We employ meta learners \cite{K_nzel_2019}, which merge machine learning and causal inference to estimate the conditional average treatment effect (CATE).

For each item, let $X$ denote a vector of predictor variables, and $Y(i)$ denote an outcome variable where a coupon is allocated when $i=1$ and no coupon is allocated when $i=0$. The CATE, the effect of a coupon on an item with the data $X=x$, is given by
\begin{equation*}
    \pi(x) = \mathbb{E}[Y(1)-Y(0) \mid X = x].
\end{equation*}
We estimate CATE using the T-learner \cite{K_nzel_2019}, a typical meta-learner algorithm. Let $f_0(x)$ denote the estimated effect on the item $X=x$ without coupon, and $f_1(x)$ denote the estimated effect with coupon. Then, we can express the CATE by
\begin{equation*}
    \pi(x) = f_1(x) - f_0(x).
\end{equation*}

\subsection{Provider's sales experience rate} \label{sec_ser}
We define the coupon effect based on the item sale rate, which means $f_1(x)$ shows an item sale rate with a coupon. Here, we introduce a metric representing the probability of a provider experiencing at least one sale. Note that the $\pi(x)$ is an item unit effect, while we aim to pursue a provider unit effect.

\begin{figure}[t]
    \centering
    \includegraphics[width=\linewidth]{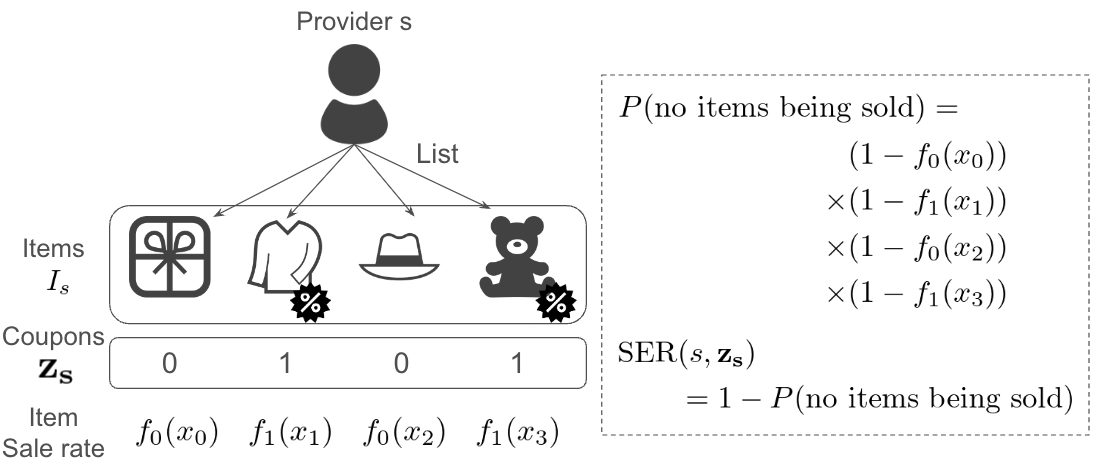}
    \caption{Provider Sales Experience Rate (SER): SER is defined using the probability of no items being sold and its complementary event probability}
    \label{fig:objective}
\end{figure}

As shown in Figure \ref{fig:objective}, we assume that a provider $s \in S$ has a set of items $I_s$ and a vector ${\bf z}_{s} \in \{0,1\}^{|I_s|}$ represents whether each item is allocated a coupon or not. Then, the provider Sales Experience Rate $\text{SER}(s, {\bf z_s})$ is then given by taking the complement of the probability that no items are sold:
\begin{align*}
    \text{SER}(s, {\bf z_s}) = 1 - \prod_{i \in I_s} \left(1-f_0(x_i)^{(1-z_s(i))} f_1(x_i)^{z_s(i)}\right),
\end{align*}
when we assume that the sales rates of items are independent of each other. By taking a difference, we get the CATE of the SER $\delta(s, {\bf z_s})$ as follows:
\begin{align*}
    \delta(s, {\bf z_s}) &= \text{SER}(s, {\bf z_{s}}) - \text{SER}(s, {\bf 0}) \\
    & = \prod_{i \in I_s} \left(1-f_0(x_i)\right) - \prod_{i \in I_s} \left(1-f_0(x_i)^{(1-z_s(i))} f_1(x_i)^{z_s(i)}\right).
\end{align*}
Due to the nature of SER's formulation, if a provider holds a large number of items or even just a few items with a high sales rate, the baseline of SER becomes high, resulting in a low CATE due to its lack of room for further uplift.

\subsection{Coupon allocation with an uplift framework} \label{coupon_alloc}
When distributing $N$ coupons, we aim to maximize the uplift in SER across all providers to maximize the total number of successful providers. The coupon allocation problem is then given by
\begin{align*}
& \underset{z_s(i)~(s \in S, i \in I_s)} {\text{maximize}} && \sum_{s \in S} \delta(s, {\bf z_s}) \\
&\text{subject to} && \sum_{s \in S} \sum_{i \in I_s} z_s(i) = N &&& \\
& && z_s(i) \in \left\{ 0,1 \right\} &&& (\forall s \in S, \forall i \in I_s).
\end{align*}
Note that here, we removed low-quality items in advance so that we can ensure the minimum quality of items with a price adjustment through coupons. Here, we define low-quality items as those that are challenging to sell even with price reductions. If there is no likelihood of sales improvement even with price reduction via coupons, we interpret this as indicating underlying issues with the items beyond pricing, such as their title, item photo, etc. As a result, we should avoid making active price adjustments through coupons since price is not the primary factor. In that scenario, we need to identify issues besides pricing and propose suggestions for improvement to the providers. In the numerical experiments section, we observe that data preprocessing for item quality assurance plays an important role.

Furthermore, given the nature of the objective function, we refrain from intervening with coupons for providers who list a substantial number of items and demonstrate strong sales performance, which raises concerns about the fairness of intervention. While the primary focus of this paper is on fairness in the outcomes of sales experiences rather than intervention fairness, the proposed objective function is deemed suitable given its aim toward outcome fairness. In situations where the fairness of intervention must be taken into account, one could include it as a constraint or explore alternative strategies.

\subsection{Integer linear programming transformation}\label{linearprob}
The uplift model with SER has a non-linearity on the shape of the objective function $\delta(s,{\bf z_s})$. By listing out all possible patterns of coupon allocation ${\bf z_s}$, the problem can be reformulated as integer linear programming.

Let $M_s = \{1,2,\ldots,2^{|I_s|}\}$ denote the index set corresponding to a set of 0-1 assignment for ${\bf z_s}$ (i.e., a set of all patterns of coupon allocation). For each $t \in M_s$, let ${\bf z_s}(t)$ denote the $t$-th 0-1 assignment. Let $w_s(t)$ denote the binary variable that takes one if a coupon allocation pattern $t \in M_s$ is chosen and zero otherwise, and $n_s(t)$ represent the number of allocated coupons corresponding to provider $s \in S$ and an allocation pattern $t \in M_s$.  By computing $\delta_s(t) = \delta(s, {z_s(t)})$ a priori and treating it as a constant value, the integer linear programming is then given as follows:
\begin{align*}
& \underset{w_s(t)~(s \in S, t \in M_s)} {\text{maximize}} && \sum_{s\in S} \sum_{t \in M_s} \delta_s(t) w_s(t) \\
&\text{subject to} && \sum_{s\in S} \sum_{t \in M_s} n_{s}(t)w_s(t) = N &&& \\
& && \sum_{t \in M_s}w_s(t) = 1 &&& (\forall s \in S) \\
& && w_s(t) \in \{0,1\} &&& (\forall s \in S, \forall t \in M_s). 
\end{align*}
Since the number of combinations grows exponentially with the number of items held by providers, we reduce the number of combinations from $2^{|I_s|}$ to $|I_s|$ by imposing a constraint to intervene in descending order of the items' sales rate uplift of the items $\pi(x)$ for each provider in the numerical experiments. For example, if the intervention is limited to only one coupon per user, assigning the coupon to the item with the highest uplift would result in the highest overall uplift. Similarly, with two coupons available, targeting the top two performing items with the highest uplift could be advantageous, rather than selecting other combinations such as the first and third highest uplift items. Therefore, such a coupon allocation constraint may have minimal impact on the overall effectiveness of uplift.

\section{Numerical experiments}
In this section, we evaluate the performance of the proposed uplift model of SER for a coupon allocation problem using an online dataset of randomized control trials.

\subsection{Dataset}
We use real-world data of digital coupons on an online marketplace app operated by Mercari Inc., Japan’s largest C2C marketplace service. Our dataset consists of multiple iterations conducted on different weeks of February 2024, with approximately 2 million providers and several million coupons. The coupons are distributed randomly to each item of targeted providers, i.e., it is a Randomized Control Trial (RCT) among the targeted providers. Naturally, there are instances where multiple coupons are allocated to a single provider for various items. We use a single type of coupon and observe whether each item is sold within one week after the coupon is distributed.

In this paper, we use data from four iterations of experiments. We train a T-learner \cite{K_nzel_2019} with LightGBM \cite{NIPS2017_6449f44a} using data from the first two iterations. We train it using 21 features, such as recent view counts, comment numbers, item categories, etc. The remaining two iterations are used as test data. However, for computational efficiency, we perform a random sampling of 10,000 providers from each test data and calculate each performance metric (Figure \ref{fig:STR} - \ref{fig:numTR}).

\subsection{Coupon allocation strategies}
We compare our proposed method with greedy and fair methods at the item level. Additionally, we compare it with the greedy method at the provider level to verify the efficiency of our proposed method in generating more successful providers. The details of each method are as follows:

\begin{itemize}
    \item {\bf Random}: allocating coupons to items randomly until all coupons are used up.
    \item {\bf Item unit Greedy (I-Greedy)}: allocating coupons for the top $N$ items with high sales rates lift $\pi(x)$ in descending order.
    \item {\bf Provider's unit Greedy (P-Greedy)}: allocating coupons to items with the highest sales rates lift $\pi(x)$ of each provider, one at a time, then on the second round intervene in the next highest sales rates lift items, continuing recursively until $N$ items are allocated. This is corresponding to the fairness of process \cite{Wang_2023}.
    \item {\bf NSW model}: The Impact-based fairness model \cite{Saito_2022} formulates its objective function as the product of all item sales rates, incorporating the concept of Nash Social Welfare (NSW).
    \item {\bf SER model (Ours)}: The SER uplift model approach. We set an item quality threshold introduced at the section \ref{coupon_alloc} by percentiles of all items predicted sales rates:
    \begin{itemize}
        \item 0\%: no item quality limit
        \item 1\%: 1 percentile of all items' predicted sales rates
        \item 10\%: 10 percentile of all items' predicted sales rates
    \end{itemize}
\end{itemize}

The SER model is solved using the mathematical optimization solver PuLP\footnote{https://coin-or.github.io/pulp/}. Instead of solving it as a nonlinear optimization problem using commercial solvers such as Gurobi\footnote{https://www.gurobi.com/}, we solve them as integer linear programming as formulated in the section \ref{linearprob}. 

\subsection{Evaluation methodology}
We define two types of evaluation metrics. First, we define the uplift in the number of items sold, which is the focus of item-level algorithms such as I-Greedy or the NSW model. Second, we define the uplift in the number of successful providers, which is the target metric for provider-level algorithms such as P-Greedy or the SER model.

In RCT, we can obtain unbiased estimates of the metrics by including only the records in which the action taken (i.e., receiving a coupon or not) aligns with the random assignment of treatments in the RCT. Table \ref{tab:item_type} shows the types of item sets. In the appendix, we prove that the evaluation metrics to be introduced are unbiased estimators of CATE in the RCT scenarios.

\begin{table}[ht]
\centering
\small
\renewcommand{\arraystretch}{1.3}
\begin{tabular}{|c|c|c|c|c|}
    \hline
    \multicolumn{3}{|c|}{} & \multicolumn{2}{|c|}{Current Algorithm Decision} \\ \cline{4-5}
    \multicolumn{3}{|c|}{\multirow{-2}{*}{}} & w/o Coupon (0) & w/ Coupon (1) \\ \hline
    \multirow{4}{*}{\rotatebox{90}{RCT}}&\multirow{2}{*}{w/o Coupon (0)} & Sold (S) & $I^S_{00}$ & $I^S_{01}$ \\ \cline{3-5} 
    && Not Sold (N) & $I^N_{00}$ & $I^N_{01}$ \\ \cline{2-5}
    &\multirow{2}{*}{w/ Coupon (1)} & Sold (S) & $I^S_{10}$ & $I^S_{11}$ \\ \cline{3-5} 
    && Not Sold (N) & $I^N_{10}$ & $I^N_{11}$ \\ \hline
\end{tabular}
\caption{Types of Item Sets: Items can be classified into eight segments based on the assignment during the RCT and the assignment in the current algorithm and whether they have been sold.}
\label{tab:item_type}
\end{table}

\subsubsection{{\bf Uplift in the number of items sold}}
According to Table \ref{tab:item_type}, $I_{**1}$ is the set of items for which coupons are granted according to the current algorithm. By taking records consistent with RCT, the uplift in the number of items sold is then given by
\begin{align*}
    \text{Uplift-ItemsSold} &= \underbrace{\left[\frac{|I^S_{11}|}{|I^S_{11}| + |I^N_{11}|} - \frac{|I^S_{01}|}{|I^S_{01}| + |I^N_{01}|}\right]}_{\text{Uplift in item sales rate}} \cdot \frac{|I^*_{*1}|}{2}
\end{align*}
where the first item sale rate term is the percentage of items sold with coupons, and the second item sale rate term is the percentage of items sold without coupons. By multiplying the number of intervened items, we can obtain the uplift in the number of items sold.

\subsubsection{{\bf Uplift in the number of successful providers}}
\begin{figure}[t]
    \centering
    \includegraphics[width=\linewidth]{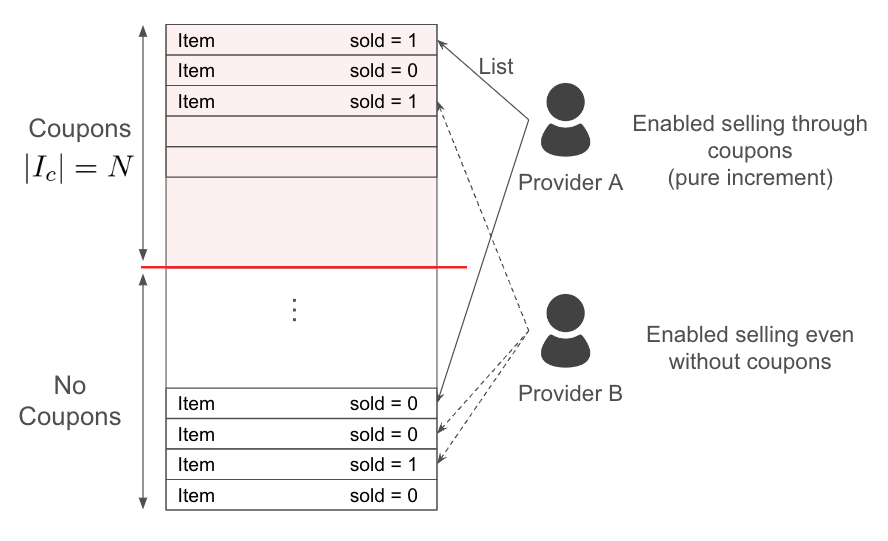}
    \caption{No items from Provider A can be sold without coupons, but one item from Provider B can be sold without coupons, so we do not need to attach coupons to make Provider B a successful provider. Coupons have resulted in an increase of successful providers through Provider A but not through Provider B.}
    \label{fig:sold_supplier_lift}
\end{figure}

Let $S^*_{**}$ be the set of corresponding providers with items $I^*_{**}$. Applying the same method as for the Uplift-ItemsSold, we can define the uplift in the number of successful providers by
\begin{align*}
    \text{Uplift-SuccessfulProviders} = \underbrace{\left[\frac{|S^{succ.}_{\text{T}}|}{|S^S_{11}| + |S^N_{11}|} - \frac{|S^{succ.}_{\text{C}}|}{|S^S_{01}| + |S^N_{01}|} \right]}_{\text{Uplift in SER}}\cdot\frac{|S^*_{*1}|}{2}
\end{align*}
where $S^{succ.}_{\text{T}}$ and  $S^{succ.}_{\text{C}}$ represent the set of successful providers for the treatment group and control group. Note that we should consider providers who can organically convert to successful providers. We aim to make a sale for at least one item among a provider's multiple items, and any further intervention should be considered wasteful (see Figure \ref{fig:sold_supplier_lift}). To consider that point, $S^{succ.}_{\text{T}}$ and $S^{succ.}_{\text{C}}$ include not only successful providers with coupons but also successful providers who could sell their items organically. By considering that point, the Uplift-SuccessfulProviders become the unbiased estimator of the CATE of the SER. The details are provided in the appendix.

\subsection{Results}
Evaluation metrics are computed using two iterations of test data.

\subsubsection{With respect to the number of items sold}
\begin{figure}[t]
    \begin{minipage}[t]{0.45\textwidth}
        \centering
        \includegraphics[width=\linewidth]{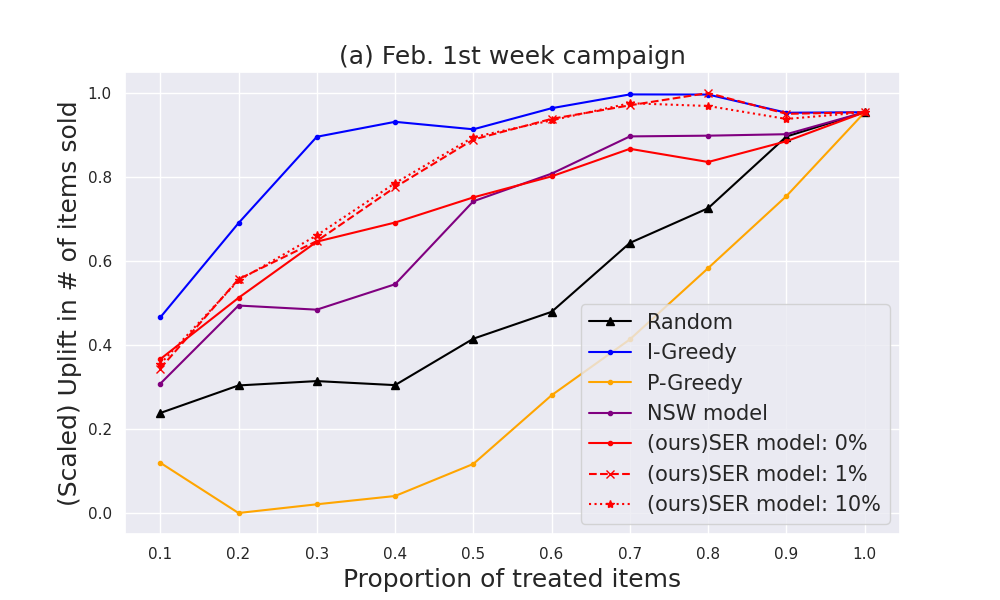}
    \end{minipage} 
    \begin{minipage}[t]{0.45\textwidth}
        \centering
        \includegraphics[width=\linewidth]{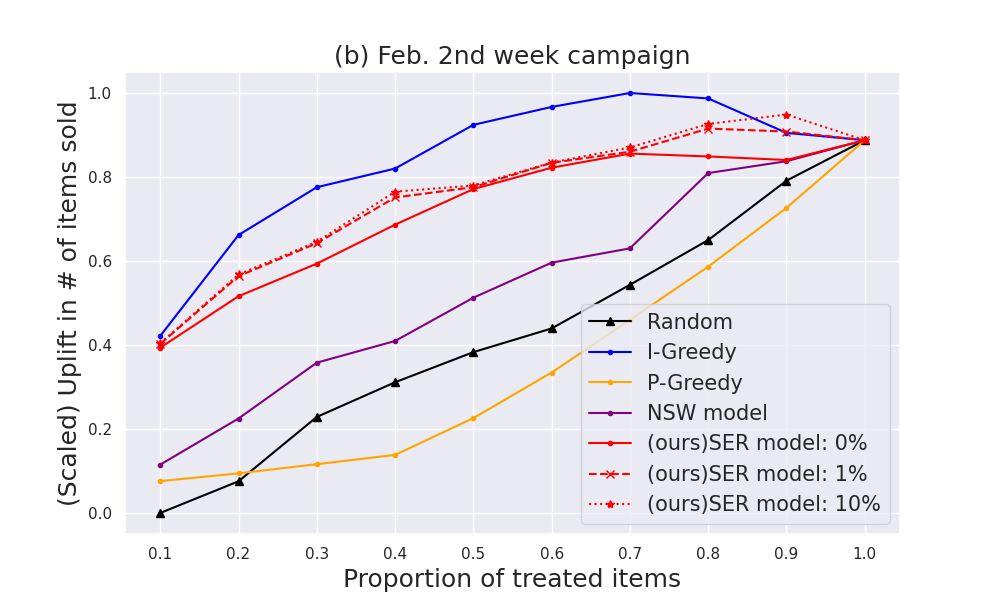}
    \end{minipage}
    \caption{Uplift in the number of items sold: For different campaigns, the item unit greedy model (I-Greedy) achieves the best lift in the number of items sold while other models seek the lift in the number of successful providers, not in the number of items sold.}
    \label{fig:STR}
\end{figure}
 Figure \ref{fig:STR} shows that the item unit greedy algorithm outperforms in terms of the uplift in the number of items sold. This superiority can be attributed to the fact that the item unit greedy algorithm is designed to maximize the item sales rate, not the successful provider rate. 

From another perspective, the NSW model tends to have a smaller uplift in the number of items sold than the SER model. This can be attributed to the NSW model's characteristic of having the objective function a product of the item sales rates, making it more susceptible to interventions on items with low sales rates. Increasing interventions on such items may not eventually lead to actual sales.

\subsubsection{With respect to the number of successful providers}
\begin{figure}[t]
    \begin{minipage}[t]{0.45\textwidth}
        \centering
        \includegraphics[width=\linewidth]{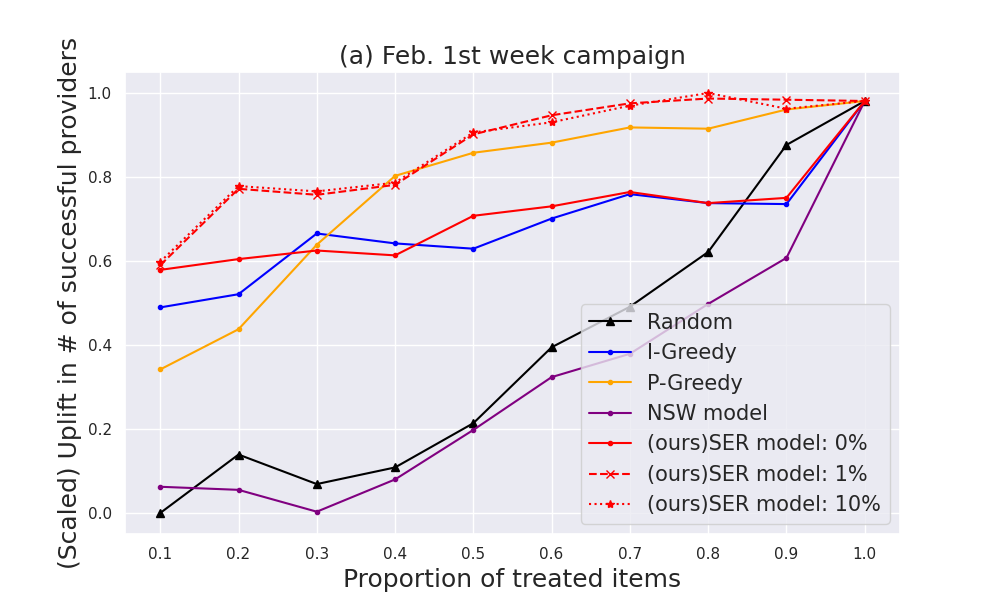}
    \end{minipage} 
    \begin{minipage}[t]{0.45\textwidth}
        \centering
        \includegraphics[width=\linewidth]{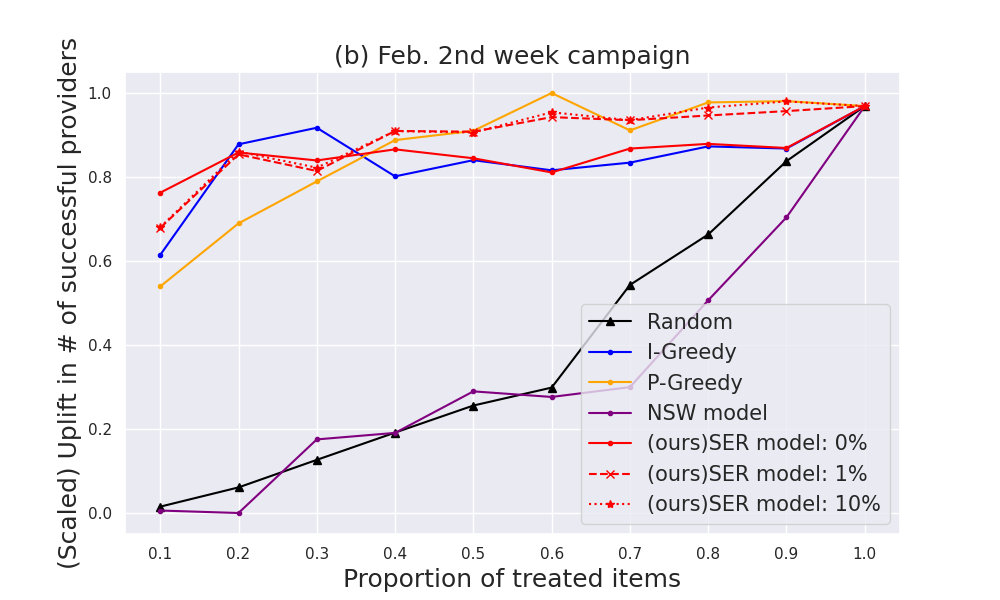}
    \end{minipage}
    \caption{Uplift in the number of successful providers: SER models with item quality assurance outperform other models focusing solely on item sale rates. Without quality assurance, the SER model's performance is close to other methods, underscoring the critical importance of this assurance preprocessing in practical applications.}
    \label{fig:SCR}
\end{figure}
Looking at the uplift in the number of successful providers, as shown in Figure \ref{fig:SCR}, SER models outperform the item unit greedy model regarding the number of successful providers. We can also see that the minimum item quality assurance plays a particularly crucial role in the SER models. Figure \ref{fig:SCR} shows that without this assurance (SER model with Q=0\%), the model only achieves performance close to the item unit greedy model. Conversely, when this assurance is included, the uplift in the number of successful providers significantly improves. Due to the nature of a two-sided market, the quality of items individuals list tends to vary considerably more than in traditional B2C businesses. Therefore, quality assurance conditions become extremely important in practical applications.

\subsubsection{With respect to the number of treated providers}
\begin{figure}[t]
    \begin{minipage}[t]{0.45\textwidth}
        \centering
        \includegraphics[width=\linewidth]{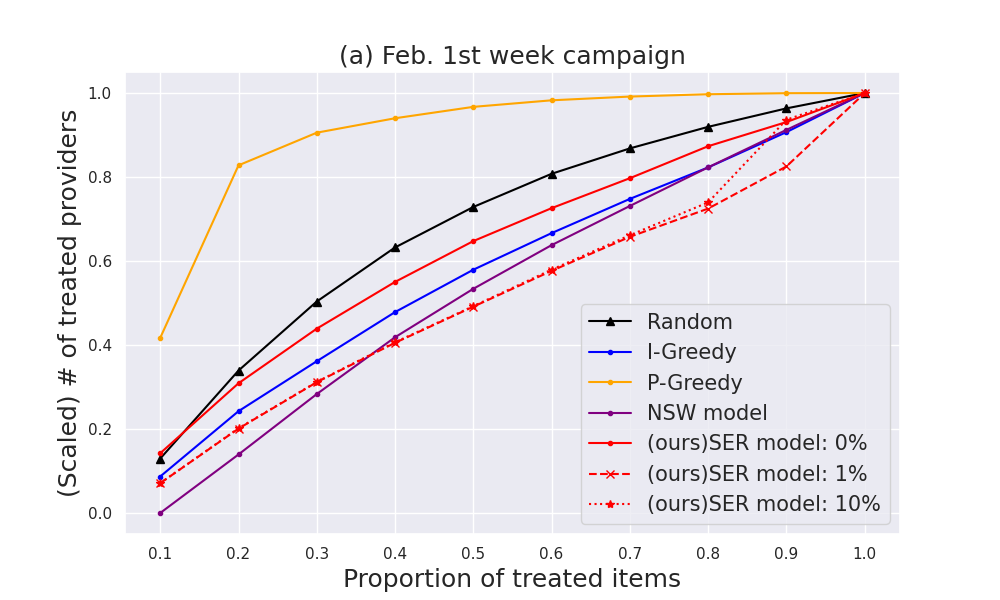}
    \end{minipage} 
    \begin{minipage}[t]{0.45\textwidth}
        \centering
        \includegraphics[width=\linewidth]{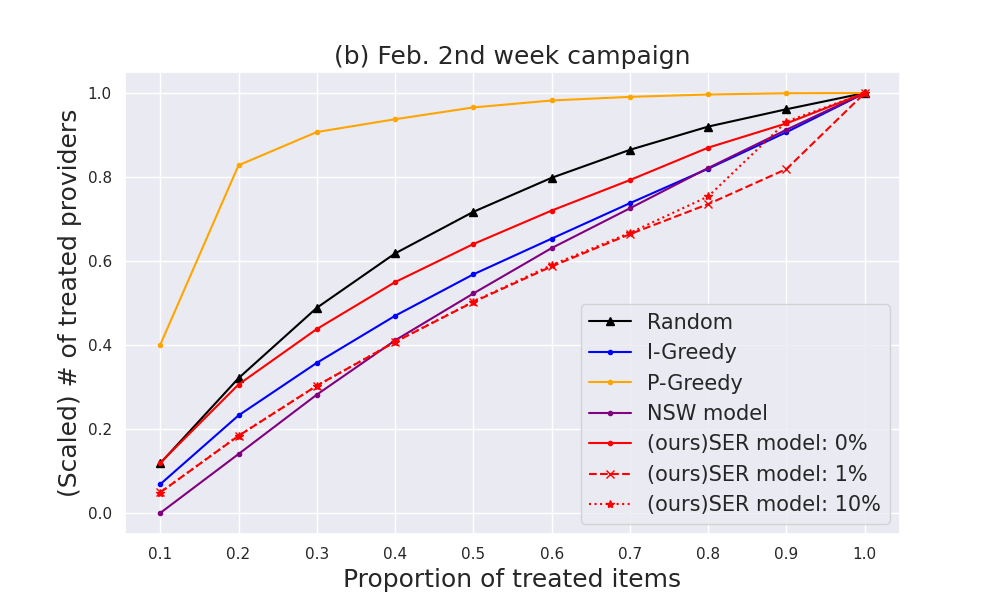}
    \end{minipage}
    \caption{The number of treated providers: Provider unit Greedy (P-Greedy) model focuses on treating more providers with the same number of coupons. On the other hand, the SER model treats fewer providers and boldly allocates coupons to the same providers to maximize their conversion rates.}
    \label{fig:numTR}
\end{figure}

Figure \ref{fig:numTR} illustrates the number of providers allocated coupons by each model when an equal number of coupons is employed. The P-Greedy focuses on treating more providers with the same number of coupons as other models. On the other hand, the SER model tends to intervene with fewer providers. This indicates the tendency of the SER model to make bold interventions by allocating multiple coupons to the same providers. As the SER model aims to maximize the overall uplift in SER, it strives for outcome fairness rather than process fairness, which means that the SER model refrains from intervening with providers who already have a high SER baseline or those with low expected uplift. Instead, the SER model boldly intervenes with providers when an uplift is anticipated.

\subsubsection{With respect to Sales Experience Rate uplift}
\begin{figure}[t]
    \begin{minipage}[t]{0.45\textwidth}
        \centering
        \includegraphics[width=\linewidth]{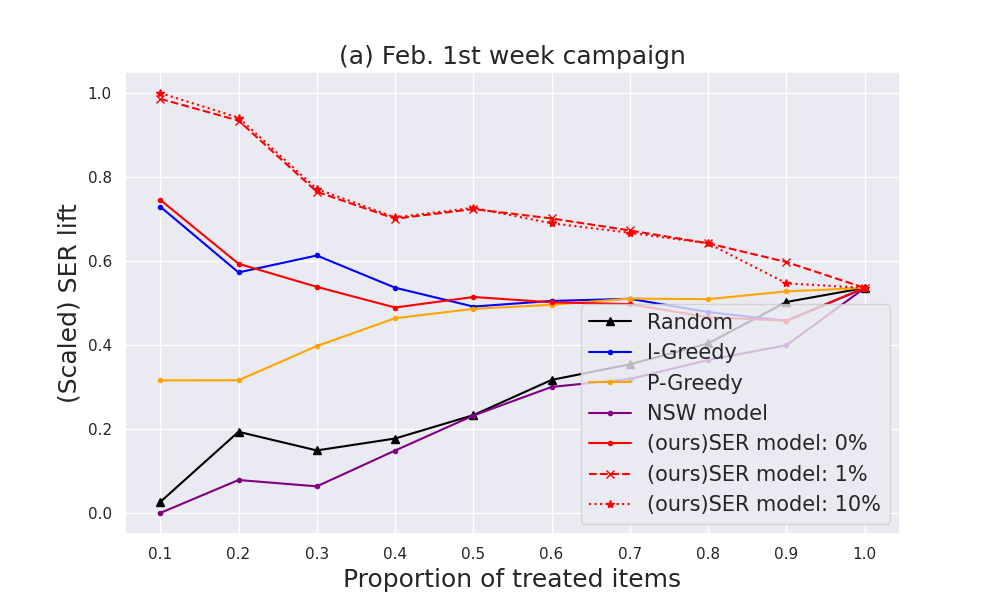}
    \end{minipage} 
    \begin{minipage}[t]{0.45\textwidth}
        \centering
        \includegraphics[width=\linewidth]{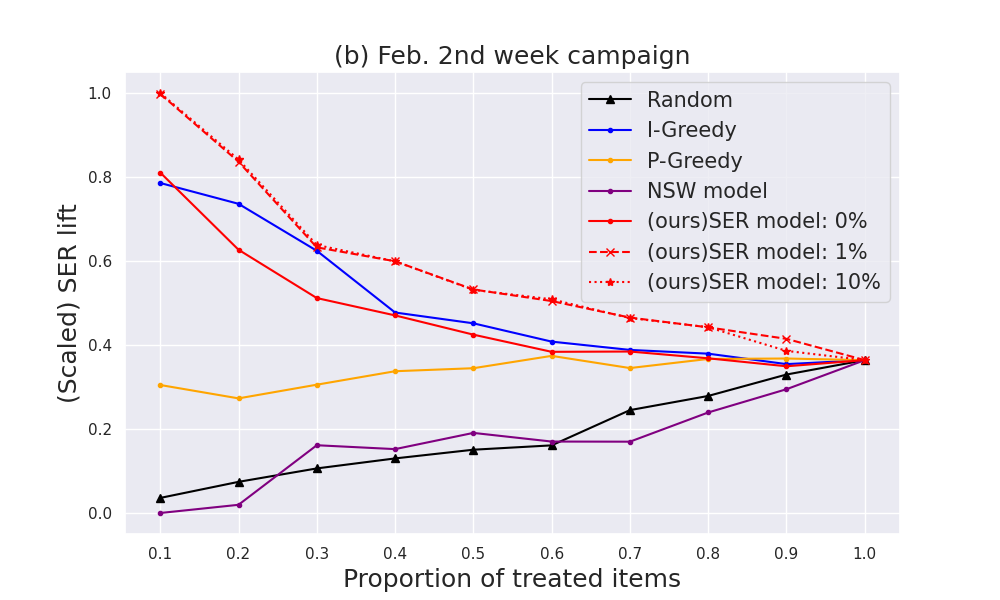}
    \end{minipage}
    \caption{SER lifts of each model: The SER models with item quality assurance show the best SER lift. This shows that item quality assurance is crucial to making a larger SER lift.}
    \label{fig:SER_lift}
\end{figure}

Figure \ref{fig:SER_lift} shows that the SER models with item quality assurance generate SER lifts most effectively. According to figures \ref{fig:numTR}-\ref{fig:SER_lift}, the P-Greedy method improves the number of successful providers by intervening with a more significant number of providers, while the SER model generates more successful providers by concentrating interventions on the part of providers as compared to P-Greedy. This suggests that the SER model, which seeks outcome fairness, can more efficiently generate successful providers than process-fair methods such as P-Greedy. The SER model can be a superior CRM strategy, particularly when aiming to offer sales experiences to a more significant number of providers to sustain the overall provider count on the platform.

\section{Conclusion}
We highlighted the importance of retaining providers in a two-sided market to sustain a competitive edge with network effects. Our proposed promotion optimization model aims to maximize the total number of successful providers who experience at least one sale, thereby contributing to provider retention. To address the practical challenge of limited access to commercial nonlinear solvers, we reformulated the optimization model as an integer linear programming problem. Additionally, we observed an enhancement in model performance by incorporating item quality assurance preprocessing to handle the spread in item quality in a two-sided marketplace.

To the best of our knowledge, no prior studies have considered whether or not having successful experiences is a fairness metric for providers. Furthermore, this paper is the first to aim to maximize the total number of successful providers within the coupon allocation framework instead of opportunity fairness based solely on rankings to boost item views. Through experiments with real data, we verified that our proposed method is more efficient in promoting successful providers than models aiming to maximize item sale rates.

\bibliographystyle{ACM-Reference-Format}
\bibliography{reference}

\appendix
\section{Asymptotic feature of evaluation metrics}
\subsection{Uplift in the number of items sold}
Assume the number of items coupon allocated $I^*_{*1}$ is large enough, 
\begin{align*}
    |I^*_{01}| = |I^*_{11}| =  \frac{|I^*_{*1}|}{2} 
\end{align*}
hold under an RCT setting due to its random allocation nature. Then, the following hold
\begin{align*}
    \mathbb{E}[\text{Uplift-ItemsSold}] &= \left\{\mathbb{E}\left[\frac{|I^S_{11}|}{|I^S_{11}|+|I^N_{11}|}\right] - \mathbb{E}\left[\frac{|I^S_{01}|}{|I^S_{01}|+|I^N_{01}|}\right] \right\} \cdot \frac{|I^*_{*1}|}{2} \\
    &= \mathbb{E}\left[|I^S_{11}|\right] -\mathbb{E}\left[|I^S_{01}|\right],
\end{align*}
where $\mathbb{E}[|I^S_{11}|]$ is the expected number of items sold with coupon allocation and $\mathbb{E}[|I^S_{01}|]$ is the expected number of items sold without coupon allocation according to the definition of item sets (see Table \ref{tab:item_type}). This is a CATE in the number of items sold by coupons.

\subsection{Uplift in the number of successful providers}
According to Table \ref{tab:item_type}, $S^*_{0*}$ represents control sellers whose items were not allocated coupons, and $S^*_{1*}$ represents treatment sellers whose items were allocated coupons at least for one item in the RCT. The uplift in the number of successful providers is then calculated by following steps:

\begin{enumerate}
    \item Extract the set of successful providers with a coupon allocation of a treatment group $S^{succ.*}_{\text{T}}$ and a control group $S^{succ.*}_{\text{C}}$ by
    \begin{align*}
        S^{succ.*}_{\text{T}} &= S^S_{11}\\
        S^{succ.*}_{\text{C}} &= S^S_{01}
    \end{align*}

    \item Extract the set of providers $S^{\text{OC}}$ who could sell their items without coupons (Organic Conversion). 
    \begin{align*}
        S^{\text{OC}}_{\text{T}} = \left\{s \in S^*_{11} \mid \exists i \in I^S_{00}(s)]\right\}\\
        S^{\text{OC}}_{\text{C}} = \left\{s \in S^*_{01} \mid \exists i \in I^S_{00}(s)]\right\}\\
    \end{align*}
    
    \item Add providers who can sell their items without coupons $S_{\text{OC}}$ to the set of successful providers $S^{succ.*}_{\text{T}}$ and $S^{succ.*}_{\text{C}}$. We aim to make a sale for at least one item among a provider's multiple items, and any further intervention should be considered wasteful (see Figure \ref{fig:sold_supplier_lift}). To consider that point, we define the set of successful providers as follows:
    \begin{align*}
        S^{succ.}_{\text{T}} &= S^{succ.*}_{\text{T}} \cup S^{\text{OC}}_{\text{T}}\\
        S^{succ.}_{\text{C}} &= S^{succ.*}_{\text{C}} \cup S^{\text{OC}}_{\text{C}}
    \end{align*}

    \item Finally, the uplift in the number of successful providers is given by
    \begin{align*}
        \text{Uplift-SuccessfulProviders} = \left[\frac{|S^{succ.}_{\text{T}}|}{|S^S_{11}| + |S^N_{11}|} - \frac{|S^{succ.}_{\text{C}}|}{|S^S_{01}| + |S^N_{01}|} \right] \cdot\frac{|S^*_{*1}|}{2}.
    \end{align*}
\end{enumerate}

When the number of items coupon allocated $I^*_{*1}$ is large enough, the number of treated providers $S^*_{*1}$ also becomes large enough. Then,
\begin{align*}
    |S^*_{01}| = |S^*_{11}| =  \frac{|S^*_{*1}|}{2} 
\end{align*}
hold under an RCT setting due to its random allocation nature. The expectation of Uplift-SuccessfulProviders is then given by
\begin{align*}
    \mathbb{E}[&\text{Uplift-SuccessfulProviders}] \\ &= \left\{\mathbb{E}\left[\frac{|S^{succ.}_{\text{T}}|}{|S^S_{11}| + |S^N_{11}|}\right] - \mathbb{E}\left[\frac{|S^{succ.}_{\text{C}}|}{|S^S_{01}| + |S^N_{01}|} \right]\right\}\cdot\frac{|S^*_{*1}|}{2} \\
    &= \mathbb{E}\left[|S^{succ.}_{\text{T}}|\right] -\mathbb{E}\left[|S^{succ.}_{\text{C}}|\right],
\end{align*}
where $\mathbb{E}[|S^{succ.}_{\text{T}}|]$ is the expected number of successful providers with coupon allocation, and $\mathbb{E}[|S^{succ.}_{\text{C}}|]$ is the expected number of successful providers without coupon allocation. According to its definition, $S^{succ.}_*$ includes successful providers who could sell their items organically, which matches the definition of Sales Experience Rate $\text{SER}(s, {\bf z_s})$ introduced in the section \ref{sec_ser}.

\end{document}